\documentclass{aa}  

\usepackage{graphicx}
\usepackage{txfonts}
\usepackage{lipsum}
\usepackage{lscape}             
\usepackage{placeins}           
 \usepackage{xcolor}                               
\usepackage{booktabs} 
\usepackage{hyperref}

\begin{document}

   \title{CosmoGen: A genetic algorithm framework for the exploration of dark energy dynamics}


   \author{D. Castel\~ao\inst{1,2}\fnmsep\thanks{\email{dmcastelao@ciencias.ulisboa.pt}}
        \and I. Tereno\inst{1,2}
        }

   \institute{Instituto de Astrof\'isica e Ci\^encias do Espa\c{c}o, Faculdade de Ci\^encias,
Universidade de Lisboa, Tapada da Ajuda, PT-1349-018 Lisboa, Portugal
            \and Departamento de F\'isica, Faculdade de Ci\^encias, Universidade de Lisboa, Edif\'icio C8, Campo Grande, PT1749-016 Lisboa, Portugal\\}

   \date{Received September 20, 2025}

 
  \abstract
  {The standard Lambda cold dark matter ($\Lambda$CDM) paradigm of the physical Universe suffers from well-known conceptual problems and is challenged by observational data. Alternative models exist in the literature, both phenomenological and physically motivated, but many of them suffer from similar or new problems.}
   {We propose a method to mechanically generate alternative models in a data-informed procedure tuned to mitigate specific problems.}
   {We implemented a computational framework, dubbed CosmoGen, based on evolutionary algorithms for symbolic regression. The evolutionary process is guided by the computation of structure formation and background cosmological quantities. As a proof-of-concept, we applied the procedure to the specific case of dark energy fluid models and asked the framework to generate models capable of alleviating the cosmological tensions $S_8$ and $H_0$.}
   {The system generated models with high fitness values, and through a Bayesian analysis of an illustrative model, we show that the model indeed alleviates the tensions, even though the Bayes factor indicates a weaker preference for $\Lambda$CDM.}
  {}

   \keywords{methods: numerical --
                dark energy --
                large-scale structure of the universe -- cosmological parameters
               }

   \maketitle

\nolinenumbers

\section{Introduction}
A quarter of a century after its discovery \citep{1998AJ....116.1009R, 1999ApJ...517..565P}, the accelerated expansion of the Universe remains a central puzzle of modern cosmology. The cosmological constant, which can be included in the Einstein equations in a natural way, has been the standard way to account for the accelerated expansion, establishing the Lambda cold dark matter ($\Lambda$CDM) paradigm. Although this paradigm has been successful in explaining current observations of the Universe, it faces significant challenges, from the interpretation of the cosmological constant as the vacuum energy density \citep{1989RvMP...61....1W} to the coincidence problem \citep{Zlatev:1998tr} to the tensions between low- and high-redshift measurements of the Hubble constant and the $S_8$ parameter \citep{2014A&A...571A..16P, 2016ApJ...826...56R, 2017MNRAS.465.1454H}. 

The theoretical investigation of dark energy alternatives to the cosmological constant as a source of accelerated expansion is a very active field of research, and there
is a long list of proposed models of various types \citep{2021CQGra..38o3001D}. Since the early days of this puzzle, a programme of observational probes of dark energy was initiated \citep{2006astro.ph..9591A}. Through measurements of background-level and large-scale structure formation observables, $\Lambda$CDM was thoroughly tested and the viability of alternative models was assessed. We are currently at the start of the fourth stage of this programme. Recent results of the first of the stage-IV cosmological surveys, performed with the Dark Energy Spectroscopic Instrument (DESI), favour a dynamical dark energy \citep{2025arXiv250314738D}. 
The first stage-IV space-based survey, ESA's \textit{Euclid} mission, started its 6-year nominal cosmological survey in 2024 \citep{2025A&A...697A...1E}, and it will be followed by the ground-based Legacy survey of Space and Time (LSST) of the \textit{Vera C. Rubin} observatory \citep{2009arXiv0912.0201L} and NASA's \textit{Nancy Grace Roman} Space Telescope \citep{2025arXiv250510574O}. These surveys will provide unprecedented datasets in terms of the number of tracers, image resolution, number of filters, and spectroscopic resolving power and will be accompanied by improvements in the handling of systematics. 

The traditional way to infer dark energy properties from data is to constrain the specific parameters of physical dark energy models against data or to constrain general features shared by various models such as the derivative of the equation of state of dark energy at a pivot point \citep{2003PhRvL..90i1301L}. However, the vast amounts of data and improved likelihoods may also be used to guide the construction of the cosmological models themselves. Today, this approach is possible thanks to the new computational techniques introduced by machine learning (ML). This is the approach followed in this work, where we create phenomenological models by generating cosmological functions according to data features encoded in the likelihood. For this, we developed a framework that integrates ML techniques with a cosmology solver and parameter space sampling methods.

In computer science, ML is a field dedicated to developing methods and algorithms capable of learning from data to describe systems or perform specific tasks \citep{Turing, CARBONELL19833}. Based on the organisation and structure of the data, ML is generally divided into two primary subfields: supervised learning and unsupervised learning \citep{2025IAUS..368...40W}. If the sample data are labelled, that is, each data point is associated with a known correct measurement, then it is a supervised learning problem. The purpose of Supervised learning is to develop a predictive model that forecasts the future behaviour of a system based on past observations. This process, also called fitting or training, involves adjusting model parameters to match the model closer to the observations. Common examples of supervised learning tasks include classification, regression, and time series modelling. If the data are unlabelled, it is an unsupervised learning problem, where the goal is to identify hidden patterns or structures within the data. Typical examples of unsupervised learning tasks include clustering and dimensionality reduction.

There are many applications of ML in physics, astronomy, and cosmology (see \citet{2019RvMP...91d5002C,2020WDMKD..10.1349F}, \citet{2019BAAS...51c..14N}, respectively, for reviews), and some of them have used symbolic regression, the ML technique we use in this work.
Symbolic regression (\citealt{Kronberger2024}) is a type of supervised learning optimised for regression problems. Today, the most used learning algorithms for symbolic regression are evolutionary algorithms inspired by principles of natural selection \citep{Goldberg, haupt}. The symbolic regression technique is capable of building functional forms by combining a set of operators, operands, and rules.
Unlike classic regression techniques, which require predetermined functional forms, symbolic regression optimises the functional form, the order of operators, and the numerical parameters. Symbolic regression enables the exploration of a range of potential equations that best fit the data. Thus, a goal of symbolic regression is to discover `physical laws' that can be analytically expressed from the data. As an example, symbolic regression has been successfully applied in rediscovering planetary motion laws using the trajectories of the Sun, planets, and major moons in the Solar System \citep{2023MLS&T...4d5002L}. 

Symbolic regression has been used in cosmology, for example, in studies of N-body simulations \citep{2021PNAS..11822038L}, baryonic feedback \citep{2023MNRAS.522.2628W}, and enhanced models of halo occupation distribution \citep{2022MNRAS.515.2733D}. More relevant for our goal, symbolic regression has also been used to generate cosmological functions \citep{2020PhRvD.101l3525A, 2023PhRvD.107j3522K, 10136815}. Most studies that use symbolic regression to generate phenomenological cosmological functions guided by data 
rely only on background-level observables. In contrast, our framework \texttt{CosmoGen} uses both  background and large-scale structure data, allowing the symbolic regression algorithms to explore the vast space of potential dark energy models. 

The new framework, \texttt{CosmoGen}, integrates supervised learning symbolic regression methods with the cosmological Boltzmann code \texttt{CLASS} \citep{2011arXiv1104.2932L, 2011JCAP...07..034B} and the Bayesian inference tool \texttt{MontePython} \citep{2013JCAP...02..001A, Brinckmann2019}.  In this framework, each candidate model proposed by the symbolic regression algorithm is implemented on the fly as a modified version of \texttt{CLASS}, treating it as a new dark energy or dark matter component. \texttt{MontePython} then performs a preliminary parameter estimation using a likelihood tuned to the objectives to be achieved, and it returns a fitness score. The score is fed back into the evolutionary loop of the symbolic regression algorithm. This is a hybrid approach between traditional sampling methods and evolutionary strategies. The evolutionary methods are tasked with discovering functional forms, whereas the traditional sampling methods are responsible for evaluating each functional form and extracting the best-fit values for its parameters.

In Sect.~\ref{ML}, we introduce the evolutionary methods used by the ML part of the framework. In Section~\ref{CG} we describe the \texttt{CosmoGen} framework, including its prototype implementation. In Sect.~\ref{models} the framework is applied to a case study, namely, the building of dark energy fluid models generated to alleviate the Hubble and $S_8$ tensions. One model from the resulting population of generated models is selected and analysed in Sect.~\ref{model}, where its parameters are constrained. We conclude our work in Sect.~\ref{conclusions}.

\section{Evolutionary methods}
\label{ML}

Symbolic regression has been integrated into a wide number of evolutionary methods. These methods can be organised into the following categories \citep{Kronberger2024}: regression-based divided in linear and non-linear methods, expression tree-based including genetic programming (GP; \citep{Koza1992}), reinforcement learning \citep{RL_book}, and transformer neural networks \citep{Vaswani2017}. There are also many variants that include several combinations of these methods, such as AI-Feyman \citep{aifeyn_0}, a physics-inspired method, and symbolic metamodel \citep{Abroshan2023}, a mathematics-inspired method.

Genetic programming (\citealt{Koza1992}) is an evolutionary algorithm widely used in symbolic regression. In evolutionary algorithms, a population of solution candidates is used in a parallel search process. This approach involves repeatedly recombining, mutating, and evaluating portions of solution candidates to gradually achieve improvements. The algorithm simulates processes observed in natural evolution, particularly the inheritance of traits from parents to offspring and selection pressure. 
GP algorithms are tree-based expressions composed of primitives, for example, binary and unary operators such as multiplication or exponential, and terminals that represent the variables and constants used to formulate a function. These are defined in what is called a `PrimitiveSet'. 

The GP algorithm proceeds as follows:
\begin{enumerate}
    \item Initialisation: A population of solution candidates (referred to as individuals) is randomly generated.
    \item Fitness evaluation: Each individual is evaluated based on a fitness function to measure its quality or how well it solves the problem.
    \item While loop or Stopping criteria: The following steps are repeated until a predefined stopping criterion is met (e.g. a maximum number of generations or satisfactory fitness level):
    \begin{enumerate}
        \item Parent selection: A subset of individuals is chosen to be parents, typically based on their fitness values.
        \item Recombination: New individuals (children) are generated from the selected parents through recombination or crossover, creating variations in the population.
        \item Fitness evaluation of children: Each child is evaluated to determine its fitness.
        \item Replacement: Based on their fitness values, individuals from the current population and from the set of new children are selected to form the next population.
    \end{enumerate}
    \item Return the best individual: The algorithm outputs the best-performing individual from the population.
\end{enumerate}

In this work we use GP as the evolutionary algorithm because of its simplicity in implementation. However, the method shows great inefficiency in the production of candidates. This is due to the fact that tree-based methods for a certain space of genotypes and for a specific tree length limit can generate multiple expressions that represent the same function \citep{10.1007/978-3-031-70055-2_17}. Exhaustive symbolic regression \citep{10136815} is an example of an alternative method that could be implemented on future versions of CosmoGen that addresses the inefficiency problems of GP in the context of cosmological model generation.

Evolutionary algorithms also need to handle a second problem, namely, overfitting. For this, the algorithms in general aim for a balance between an accurate fit and simple equations. Indeed, while more complex functions can achieve a higher accuracy on specific datasets, they tend to generalise poorly due to overfitting. Hence, limiting the model complexity is an important point since simpler and interpretable expressions, though not always the most accurate, frequently exhibit a more reliable extrapolation behaviour. Achieving this balance between accuracy and simplicity to derive the `best' equation for a given dataset requires a procedure for managing these two aspects. This can be done by including a penalty for complex models in the fitness evaluation or, alternatively, (as done in this work) by applying an evolutionary strategy with a large initial population and a population of mixed generations at each iteration, which prevents a fast increase of model complexity.

\section{CosmoGen framework}
\label{CG}

The \texttt{CosmoGen} framework contains two primary components: the ML component and the cosmological component. The ML component consists of a symbolic regression method, which is implemented via the Distributed Evolutionary Algorithms in Python (DEAP) package.\footnote{https://pypi.org/project/deap/} The DEAP is a framework for implementing evolutionary algorithms, including but not limited to GP. The cosmological component is linked to a modified version of \texttt{CLASS}.

The ML component focuses on building the structure of the cosmological equation, i.e. the functional form of the candidates. The choice of method used to generate a population of candidates can be adjusted with ease. 
Currently, the framework uses a standard tree-based GP from \texttt{DEAP}, with minimal modifications.
The cosmological component evaluates the candidates for physical viability, integrating \texttt{CLASS} computations and \texttt{MontePython} sampling methods. This second component is integrated in the GP loop, acting on points two and three of the procedure outlined in Sect.~\ref{ML}.

The physical interpretation of each individual depends on the desired case study and is determined by how it is integrated into the modified version of \texttt{CLASS}. Our framework provides a wide range of possibilities for exploration. It allows the user to consider any new cosmological component for which the relevant cosmological equations for background and perturbations can be written in a generalised form. 

Fixing the generalised equations provides the Ansatz for the new cosmological component. \texttt{CosmoGen} generates new candidates considering the specific Ansatz. 
Three modified versions of \texttt{CLASS} can be included in CosmoGen, allowing for three Ansaetze: one designed for the potential of scalar fields (which can represent either scalar field dark matter or dark energy), one for a cosmological fluid (which may be dark matter, dark energy, or unified dark matter-energy), and another to explore models with couplings between dark matter and dark energy. Only the cosmological fluid case is fully implemented in the current version of CosmoGen.

In the case study presented here, we consider a dark energy fluid $\rho_{\rm DE} (a)$ with no perturbations that will replace the cosmological constant in the standard $\Lambda$CDM model.
This case is implemented by adding a new dark energy fluid instance to \texttt{CLASS} (and setting  $\Omega_{\Lambda} = 0$.). 

\begin{figure}[t!]
    \centering
    \includegraphics[width=\hsize]{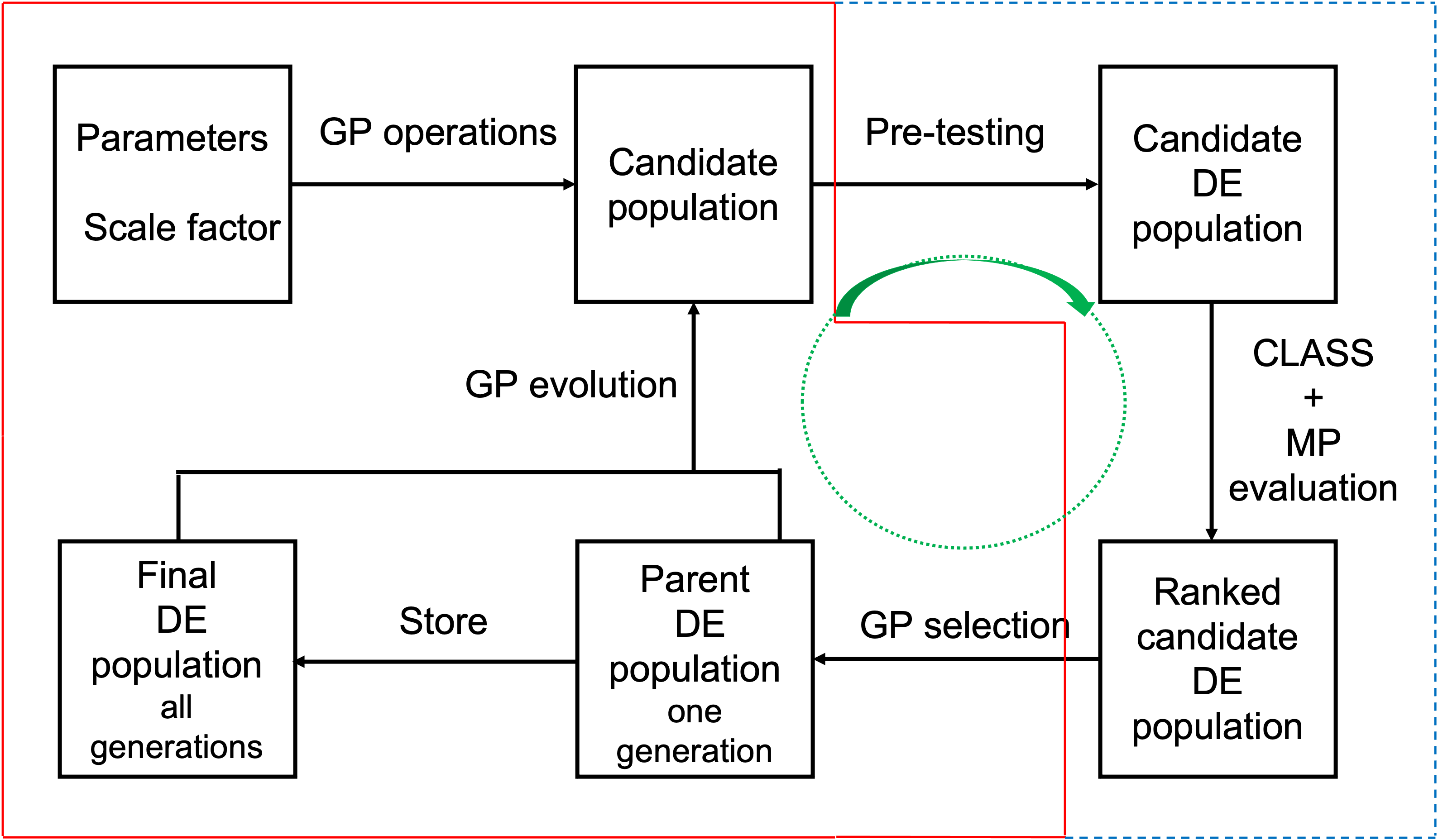}
    \caption{Flowchart showing the sequence of steps of a \texttt{CosmoGen} run. The ML component is enclosed in the solid red border, and the cosmological component is within the dashed blue border. The sequence of steps moves through the two components in a loop. The terms `GP' and `MP' stand for GP and MontePython, respectively. See text for a full description.}
    \label{fig:CGdiagram}
\end{figure}

To run \texttt{CosmoGen}, the following components are required:
\begin{itemize}
    \item The \texttt{DEAP} package, which provides the evolutionary algorithms used for symbolic regression, along with other standard Python packages. These can be easily installed through \texttt{Pip} or \texttt{Conda}.\\
    \item A compiled version of \texttt{CLASS}, which is used for cosmological computations.\\
    \item The \texttt{MontePython} code, which is employed for the optimisation of the coefficients using Markov chain Monte Carlo (MCMC) methods.
\end{itemize}
\texttt{CosmoGen} contains an input file that defines the GP configuration for the simulation as well as the paths for \texttt{CLASS}, \texttt{MontePython}, and the simulation output. The code runs following the procedure described in 
Sects.~\ref{CGstep1} through ~\ref{CGstep5}, which is summarised in Fig.~\ref{fig:CGdiagram}.

\subsection{Initial population}
\label{CGstep1}

The algorithm begins by randomly generating an initial population of candidates. This is done by applying a predefined set of operations (e.g. addition, multiplication, sine, logarithm, negation) to a predefined number of variables (e.g. the scale factor) and coefficients (free parameters to allow different coefficients on the various terms of the functional forms generated).

\subsection{Pre-testing}
\label{CGstep2}

The evolutionary process then starts from this first generation and continues in an iterative loop. Each step of the loop corresponds to one generation. In each iteration, the population passes through a pre-testing procedure that filters the individuals that have the potential to be a dark energy model. The pre-testing consists of the following steps applied to each individual of the population.

Simplification: The individuals are simplified using both \texttt{SymPy} and customised simplification methods (\texttt{smart\_simplify}) that perform a permutation check and re-parametrisation of constants. This step ensures that the expressions are in their most reduced form for further processing.

Energy density: After the simplification step, the functional forms are transformed into an energy density expression by multiplying the original functional forms by an amplitude parameter A representing an energy density parameter.

Uniqueness check: The framework checks if the generated individuals have been previously generated by comparing it to a history of past equations. If the expression is not unique, the individual is rejected.

Calculation of derivatives: The first and second order derivatives of the energy density are computed. These derivatives are needed to calculate the equation of state $w(a)$ using the continuity equation,
    \begin{equation}
        w=-\frac{a\,\rho'}{3\,\rho}-1,
        \label{derivw}
    \end{equation}
    and the adiabatic speed of sound, $c_s^2$, using its derivative,
     \begin{equation}
        c_S^2=\frac{dp}{d\rho}=\frac{a\,\rho''}{3\,\rho'}-\frac{4}{3},
        \label{derivcs}
    \end{equation}
where the derivatives are taken with respect to conformal time.
    
Infinities and complex number check: The free parameters of the candidate equation are not assigned specific values at this point.
To verify the validity of the expressions, it is necessary to test the free parameters across a range of values. We selected three indicative values to probe their range: a value close to zero, another `mid-range' value close to unity, and a large value. For each value, we evaluated $\rho(a)$, $w(a)$, and $c_s^2(a)$ to ensure that no infinities nor any complex values were produced. 
This is critical, as complex values are not physically meaningful in this context. If no viable parameter values are found, the model is discarded.

Free parameters priors:  A penalty value is computed based on how far the equation of state $w(a)$ computed for the selected free parameters values is from a pre-defined target range. A similar computation is made for $c_s^2$. For each model, the value of the free parameters that yield the lowest penalty value is stored for future indication of the region-of-interest of the parameters. 

\subsection{Evaluation}
\label{CGstep3}

At this point, we had a subset of the population identified as possible dark energy models. The next step was the evaluation procedure, which is a core component of \texttt{CosmoGen}. The evaluation step is responsible for evaluating the fitness of each individual and consists of several steps, applied in turn for each surviving individual. We describe the process in the following.

Conversion: The energy density expression and its derivatives are mechanically updated in the format required by the \texttt{CLASS} code using a built-in library.
        
File modification: The background and input source files from \texttt{CLASS} are modified to include the new dark energy fluid. Input files for both \texttt{CLASS} and \texttt{MontePython} are also updated.
        
Compilation: The modified version of \texttt{CLASS} is compiled. If the compilation is successful, \texttt{CLASS} performs a single test run using the best free parameter value from the pre-testing phase.
        
MCMC evaluation: \texttt{MontePython} runs an MCMC that samples the parameter space of the free parameters plus additional cosmological parameters. The free parameters are assigned priors according to the best values found in pre-testing. The choice of likelihood is a key aspect of the procedure. This is the point where the data guide the generation of the population of models. The likelihood is selected according to the problem that is to be addressed (e.g. we may want models that alleviate cosmological tensions, models that are optimised for a given combination of datasets, or models optimised for a certain measured feature). The MCMC is run for each model that passes the pre-testing, and hence this is a computing intensive step. 
        
Fitness value: The best $\chi^2$ value from the MCMC is returned as the fitness value to be used in the subsequent evolutionary selection step. 

\subsection{Selection}
\label{CGstep4}

At the end of the evaluation, each model has a fitness value given by the best-fit $\chi^2$ obtained for the chosen likelihood and resulting from the short chains. We note that this is not necessarily the minimum $\chi^2$ that could be achieved with longer chains or a different parameter space. 

Given the uncertainty associated with these best fits, the next step involves selecting a subset of the 
the best-performing solutions. The selection process is 
implemented through a tournament method. 
In this approach, a set of the best individuals is randomly selected to compete. The winners of the tournaments constitute the final selection of the present generation. They are stored as part of the final population, joining the winners of all past generations.

\subsection{Evolutionary strategy}
\label{CGstep5}

At the end of each iteration, the resulting final population (consisting of the survivors from the current and past generations selection processes) is taken as `parents' and used to build the next generation. This mixed generation population serves as the basis for the generation of a new candidate population (the `children'). The process is implemented applying the so-called ($\mu$+$\lambda$) evolutionary strategy. The parameters of the strategy are defined as follows:

\begin{itemize}
    \item $\mu$: The number of parents randomly selected from the pool.
    \item $\lambda$: The size of the target population.
\end{itemize}
We note that $\lambda/\mu$ is the number of offspring generated by each selected parent.

The parents are combined through the evolutionary operations mate and mutate, constrained to the chosen values of $\mu$ and $\lambda$, to build the next generation of the population. In our implementation of the evolutionary strategy, we considered an initial population (first generation)  larger than $\lambda$, the size used in subsequent generations.  
Starting with a larger initial population guarantees a large number of first-generation parents in the pool and prevents a fast increase of the complexity of the models through the generations.

\section{CosmoGen run: Cosmological tensions}
\label{models}

We now focus on a proof-of-concept application of the full procedure of the \texttt{CosmoGen} framework. The main steps of this procedure are described below.

\subsection{Setup}

In this application, we asked \texttt{CosmoGen} to generate a population of dark energy models capable of alleviating the $S_8$ and $H_0$ tensions. We restricted the population to unperturbed dark energy fluids.

\subsubsection{Dark energy density}
The general equation for the normalised energy density of the dark energy component is set to the following form:
\begin{equation}
\Omega_{\rm DE}(a) = \frac{A}{f(a; D)}.
\label{DEmodel}
\end{equation}
Here, $A$ is a normalisation parameter of the dark energy density, $f(a;D)$ is the function to be generated by CosmoGen, $a$ is the scale factor, and $D$ is a free parameter.

By replacing the cosmological constant by this new dark energy component, we can calculate the current values of the normalised dark energy density from the closure relation, 
\begin{equation}
    \Omega_{\rm DE,0} = 1 - \sum \Omega_i - \Omega_{\rm K},
\end{equation}
with the index i representing all other standard components of the model and K representing the curvature. From $\Omega_{\rm DE,0}$, we  calculated the parameter $A$ as
\begin{equation}
    A = f(a=1, D) \,\Omega_{\rm DE,0}.
\end{equation}
This procedure allows us to normalise the function generated by \texttt{CosmoGen} to expected values of the energy density. Equation~\eqref{DEmodel} can thus be rewritten as
\begin{equation}
    \Omega_{\rm DE}(a) = \frac{f(a=1; D) \,\Omega_{\rm DE,0}}{f(a; D)}.
\end{equation}

\begin{table}[t!]
        \centering
        \caption{Parameters set for the evolutionary algorithm.}
        \begin{tabular}{ll}
            \toprule
            GP parameter & Value\\
            \midrule
            Operation & Add, Sub, Mul, Pow, \\
            & Div, Inv, Exp, Log, Neg\\
            Initial population size & 4096 \\
            Generations & 8\\
            $\mu$ & 128\\
            $\lambda$ & 512\\
            Mate probability& 0.5\\
            Mutate probability& 0.5\\
            Max. first gen. complexity& 7\\
            Max. complexity increase& 2\\
        \bottomrule
        \end{tabular}
        \label{tab:GPparams}
    \end{table}

\subsubsection{Evolutionary parameters}

The population is generated by a base set of operations (addition, subtraction, multiplication, power, division, inversion, exponential, logarithm, negation) applied to Eq.~(\ref{DEmodel}). Each individual of the population is thus characterised by a certain functional form -- function of the scale factor $a$ and the free parameter $D$ -- and by the derived parameter $A$.

We needed to set values for a number of parameters of the evolutionary algorithm. These included the size of the initial population; the number of generations, which is the ending condition for the loop; the number of parents, $\mu$; and the size of the populations of each generation (except the initial one), $\lambda$. We can also define the maximum complexity of the initial population, the maximum complexity increase from the generation to generation (small values favouring simple functional forms), and the tournament size (the number of individuals that compete for a single spot). The parameter values we chose are given in Table~\ref{tab:GPparams}.

\subsubsection{Pre-testing}

To set up the pre-testing, we needed to define the values of the free parameter D to use in the required computations of the models. We defined three values: 0.05, 0.95, and 1000.

We set the target range for the equation-of-state evaluation to $[-1.5,0]$. We computed $w(a)$ in ten logarithmic points of the scale factor between $10^{-5}$ and 1. For each point where the value is outside the range, the model was assigned a penalty given by the difference of the $w$ value to $-1$. The penalties are cumulative. In addition, very large penalties are assigned if $w(a)$ becomes non-real or numerically undefined. In practice, parameter values yielding a total penalty above a large threshold are rejected, and thus models producing an unphysical behaviour (e.g. extremely large or divergent values of w) are effectively excluded. The same procedure was applied to the sound speed with a penalty given by the difference of $c_s^2$ to zero. The value of D that leads to a smaller penalty is the one used in the evaluation step.

\subsubsection{Integration with CLASS}
We set the dark energy fluid as a replacer for the cosmological constant in the relevant locations in \texttt{CLASS}. For example, the Friedmann equation became
\begin{equation}
    H^2(a) = H_0^2\left[\Omega_m \left(\frac{a_0}{a}\right)^3 + \Omega_r \left(\frac{a_0}{a}\right)^4 + \Omega_{\rm DE}(a)\right],
\end{equation}
and we considered the dark energy fluid to be homogeneous.

We used the standard CLASS settings that use an adaptive integration that works for a wide range of models. Models with rapid oscillations in the power spectrum are an example of models where the integrations should require increased numerical precision, but we kept the standard settings to limit the computational cost. However, such models are excluded in the pre-testing phase through the sound speed threshold. 

\subsubsection{Integration with MontePython}

The evaluation step computes a fitness value for each individual. This is done by running MCMC chains with \texttt{MontePython} using the Metropolis-Hastings algorithm. The chains sample a parameter space evaluating, for each individual, the likelihoods of parameter vectors with respect to data.

As stated earlier, in this proof-of-concept, our goal is to use a dataset that will produce better fitness values for individuals that are capable of alleviating the cosmological tensions. Cosmological tensions refer to the fact that constraints on some of the cosmological parameters are significantly different when determined using high-redshift data, namely CMB data, and lower-redshift data, namely Supernova Type Ia data probing properties of the homogeneous Universe and weak gravitational lensing (WL) data probing the inhomogeneity of the Universe. The Hubble tension \citep{2024ARA&A..62..287V}  is the most significant of these, with a more than 4$\sigma$ difference between the  $H_0$ Planck 2018 \citep{2020A&A...641A...6P} and supernova Ia SH0ES 2022 \citep{2022ApJ...934L...7R} measurements. The $S_8$ tension shows a 2$\sigma$ difference between the $S_8=\sigma_8\,\sqrt{\Omega_m/0.3}$ measurements of \texttt{Planck} \citep{2020A&A...641A...6P} and KiDS-1000 \citep{KiDS-1000}. However, the subsequent KiDS-legacy analysis, with better handling of WL systematics, has reduced the $S_8$ tension to 0.73$\sigma$ \citep{kids-legacy}.
In pursuit of our goal, we evaluated the models using the Planck 2018 CMB temperature and polarisation likelihoods Planck\_high\_l\_TTTEEE\_lite + Planck\_low\_l\_EE + 
Planck\_low\_l\_TT combined with a Gaussian prior for $H_0$ from the SH0ES 2022 estimate and a Gaussian prior for $S_8$ from the KiDS-1000 estimate. 

We sampled a 3D parameter space consisting of the parameter $D$ (with a flat prior around the value determined in the pre-testing for each individual), $h$, and $\omega_{CDM}$. The remaining cosmological and nuisance parameters were fixed to the mean values of the Planck 2018 $\Lambda$CDM estimates. For each point in the 3D space, $S_8$ was derived from the other parameters. The cosmological functions used in the likelihood are the CMB power spectra computed with \texttt{CLASS}. We focused on dark energy models with only one free parameter in order to avoid a penalisation from information criteria when comparing our findings with $\Lambda$CDM. As shown below, this choice produced competitive models, fulfilling the goal of the proof-of-concept.

\begin{figure}[t!]
    \centering
    \includegraphics[width=0.65\hsize]{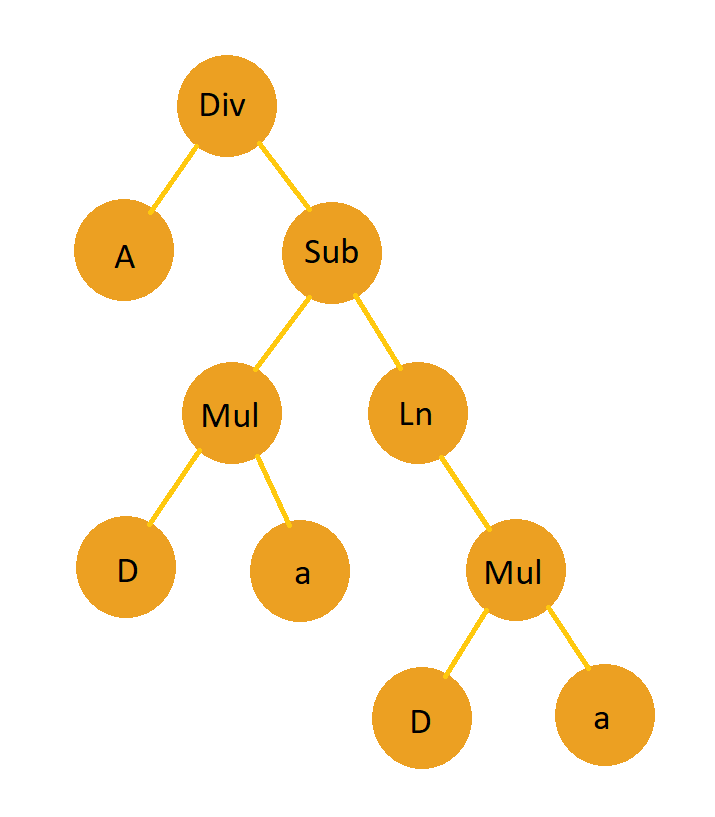}
    \caption{Representation of the function $A/\left[Da - ln(Da)\right]$ of complexity ten as a tree generated from the GP algorithm.}
    \label{fig:tree of GP}
\end{figure}

\subsubsection{Complexity}

For each individual of the final population a value of complexity is computed. By representing the functional forms as trees, one can define in a natural way a measure of their complexities as the number of nodes in the tree. 
The combination of two variables by an operator (multiplication, addition, subtraction, division, power) has a complexity of three. The application of an operator to a variable or combination of variables (logarithm, exponential) has a complexity of two. Similarly,  the combination of one variable and a constant by an operator (e.g. negation, inversion, square, cube, duplication) has a complexity of two. When counting the nodes in the example shown in Fig.~\ref{fig:tree of GP} for the model $A/\left[Da - ln(Da)\right]$, which we discuss in Sect.~\ref{model}, shows this functional form has a complexity of ten. The complexity is used to compute a merit score to each elements of the final population.

\subsection{Results}

One run of \texttt{CosmoGen} produces a final population of $o(10^2)$ models. We considered 512 models per generation throughout the search that added to the initial population size, which corresponds to a total of 8192 models involved in the full run. Each model was evaluated with the three-parameter Metropolis-Hastings MCMC. Due to the computational cost, we ran short chains and only one per model. We made preliminary tests of the step size and chain length and concluded that, in general, the chains started to stall when sampling around the posterior maximum, after less than 1000 points. This prompted us to set the chain length to 1200 points. With such a limit, not all tested models reached convergence, and some high fitness models may have been missed, biasing the sample towards lower fitness models. However, our goal is not be exhaustive but to find a small number of high fitness models to be further tested with better accuracy. With this setup, a typical run of the full process in 8 CPUs takes 240 hours.

\begin{table*}[t!]
    \centering
        \centering
        \caption{Twenty best scoring models generated by the \texttt{CosmoGen} run.}
        \begin{tabular}{cccccccc}
            \toprule
            Model & Score & $\chi_{min}^2$ & Complexity & $\log_{10}(D)$ &h & $\Omega_{\rm m}$ & $S_8$\\
            \midrule
             $\frac{A}{D^{3} a^{3} - \ln{\left(a \right)}}$ & 637.17 & 627.17 & 10 & -0.130 & 0.754 &0.247 &0.767\\
            $\frac{A}{D a^{a} - a}$ & 638.99 & 629.99 & 9 &0.088 &0.741 &0.256 &0.772\\
            $\frac{A}{D a^{a^{D}} - a}$ & 640.87 & 629.87 & 11 &0.099 &0.744 &0.254 &0.771\\
            $\frac{A}{D^{2} a^{2} - \ln{\left(a \right)}}$ & 642.03 & 632.03 & 10 & -0.117 & 0.730 & 0.261 & 0.770\\
            $\frac{A}{D a - \ln{\left(D a \right)}}$ & 644.25 & 634.25 & 10 & -0.351 & 0.737 & 0.257 & 0.769     \\
            $\frac{A}{D a^{D} - \ln{\left(a \right)}}$ & 645.09 & 635.09 & 10 & -0.184   & 0.765 & 0.240 & 0.768 \\
             $\frac{A}{D^{3} a^{a + 2} - \ln{\left(a \right)}}$ & 645.50 & 633.50 & 12 & -0.102 &        0.734 & 0.259   & 0.769 \\
            $-\frac{A}{a - \ln{\left(a^{D} \right)}}$ & 645.65 & 636.65 & 9 & -0.875      & 0.711 & 0.274 & 0.770  \\
            $\frac{A}{D \ln{\left(D a \right)} - a}$ & 646.91 & 636.91 & 10 & -0.330 &0.746 &    0.251 & 0.769 \\
            $\frac{A}{D^{3} a - \ln{\left(a \right)}}$ & 647.15 & 638.16 & 9 &   0.172 & 0.730 &0.262 &0.774  \\
            $\frac{A}{a^{D} - \ln{\left(a^{D} \right)}}$ & 647.38 & 637.38 & 10 &-0.122 & 0.696 &0.285 &   0.775  \\
            $\frac{A}{\left(D^{2}\right)^{a}}$ & 647.46 & 641.46 & 6 & -0.010 & 0.720 & 0.269 & 0.775 \\
            $\frac{A}{D^{a} \ln{\left(D a \right)} - a}$ & 647.84 & 635.84 & 12 & -0.306   & 0.763 & 0.241 &       0.768 \\
            $\frac{A}{D a^{D} - \ln{\left(D a \right)}}$ & 648.88 & 636.88 & 12 & 0.259 & 0.769 & 0.238 & 0.767  \\
            $\frac{A}{D \left(a^{D}\right)^{D} - \ln{\left(a \right)}}$ & 649.14 & 637.14 & 12 & -0.075 &0.736 &0.258 & 0.772 \\
            $\frac{A}{D^{2} a - \ln{\left(a \right)}}$ & 649.77 & 640.78 & 9 & 0.175 & 0.692 &   0.288 & 0.773 \\
            $\frac{A}{\left(D^{3}\right)^{a}}$ & 649.97 & 643.97 & 6 & -0.006 &       0.693 & 0.289 & 0.781 \\
            $\frac{A}{D^{2} a^{a} - \ln{\left(a \right)}}$ & 650.19 & 639.19 & 11 &  0.068 &0.683 & 0.294 & 0.774  \\
            $\frac{A}{D^{3} a - \ln{\left(D a \right)}}$ & 650.75 & 639.75 & 11 & 0.150 & 0.696 &  0.285 & 0.772  \\
              $-\frac{A}{D \left(D^{2} a - a^{a}\right)}$ & 657.09 & 644.09 & 13 & 0.053 & 0.690 & 0.290 & 0.776 \\
            \bottomrule
        \end{tabular}
        \label{tab:top_equations_1}
    \end{table*}

The 20 best models are shown in Table~\ref{tab:top_equations_1}. The functional form, score, minimum $\chi^2$,  complexity, and best-fit parameters are indicated for each model. The models are ranked by score, where the ones with the lowest scores are considered best. This score parameter is the result of penalising the models according to their size and is definited by adding the model complexity to the fitness given by the  minimum $\chi^2$ found in its  MCMC. This metric is motivated by the Akaike information criteria (AIC; \citealt{1974ITAC...19..716A}), where models are penalised according to the number of free parameters by summing a $\chi^2$ to a linear function of the number of parameters. In our case, since all models have the same number of free parameters, the penalisable overparametrisation is not the number of parameters but the complexity of the dark energy expression. Since the complexity is of the same order of the delta $\chi^2$ between models, the unweighted sum we applied is a less severe penalty than the AIC expression. Examples of penalty factors are discussed in \citet{Rissanen1978}, which introduces a score with a stronger penalisation than AIC and the novel fitness function of \citet{Kartelj2023}. The score is only used to rank the models at the end of the process and plays no role in the run.

We stress that the purpose of the short MCMC chains is to evaluate and select models. In other words, a high ranking shows that the model has a high likelihood for some combination of parameter values and is a strong candidate to be selected for the final population and additionally to become a parent. This conclusion can be made even with non-converged chains since the goal is not to estimate the posterior distribution. Therefore, this procedure does not guarantee that the model would remain among the top ranked ones after full MCMC analyses in a larger parameter space.

We selected one of the top ranked models for further investigation. In Sect.~\ref{model} we investigate its features and do a more complete MCMC analysis to verify if it indeed fulfils the goal of alleviating the tensions. For that analysis, we prefer a model with a good fitness score, low complexity, a reasonably looking functional form, and (if possible) resemblance with a known model. We selected the fifth ranked model from Table~\ref{tab:top_equations_1}.

We also made several additional runs of \texttt{CosmoGen} to compare the resulting final populations and tested the robustness of the procedure. We present the discussion of those results in Appendix \ref{tests}.

\section{Analysis of a selected model}
\label{model}

\begin{figure}[t]
        \centering
\includegraphics[width=0.8\hsize]{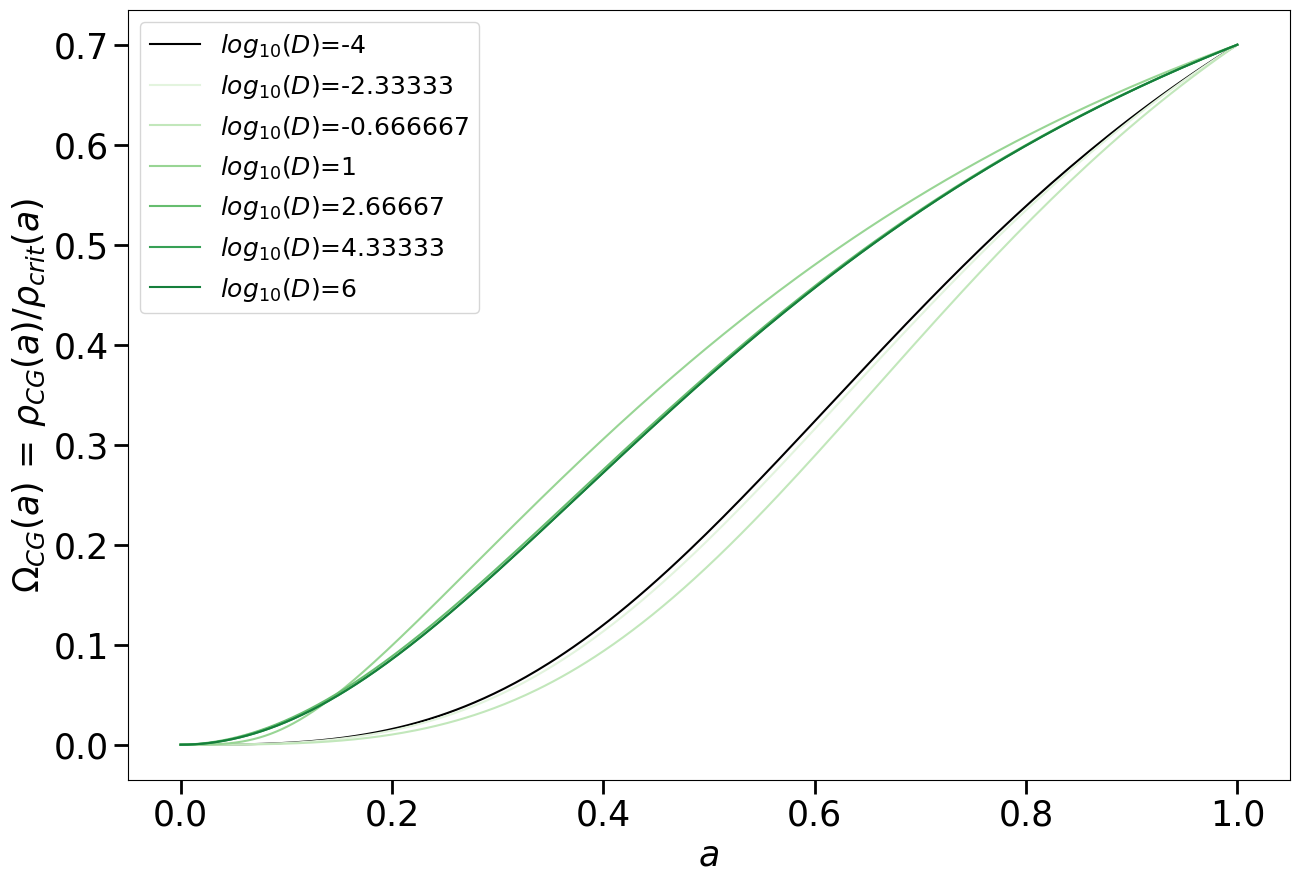}
        \caption {Normalised energy density evolution for the CG dark energy fluid for different values of its free parameter D.}
    \label{Omega_CG}
\end{figure}

\begin{figure}[t]
        \centering
\includegraphics[width=0.8\hsize]{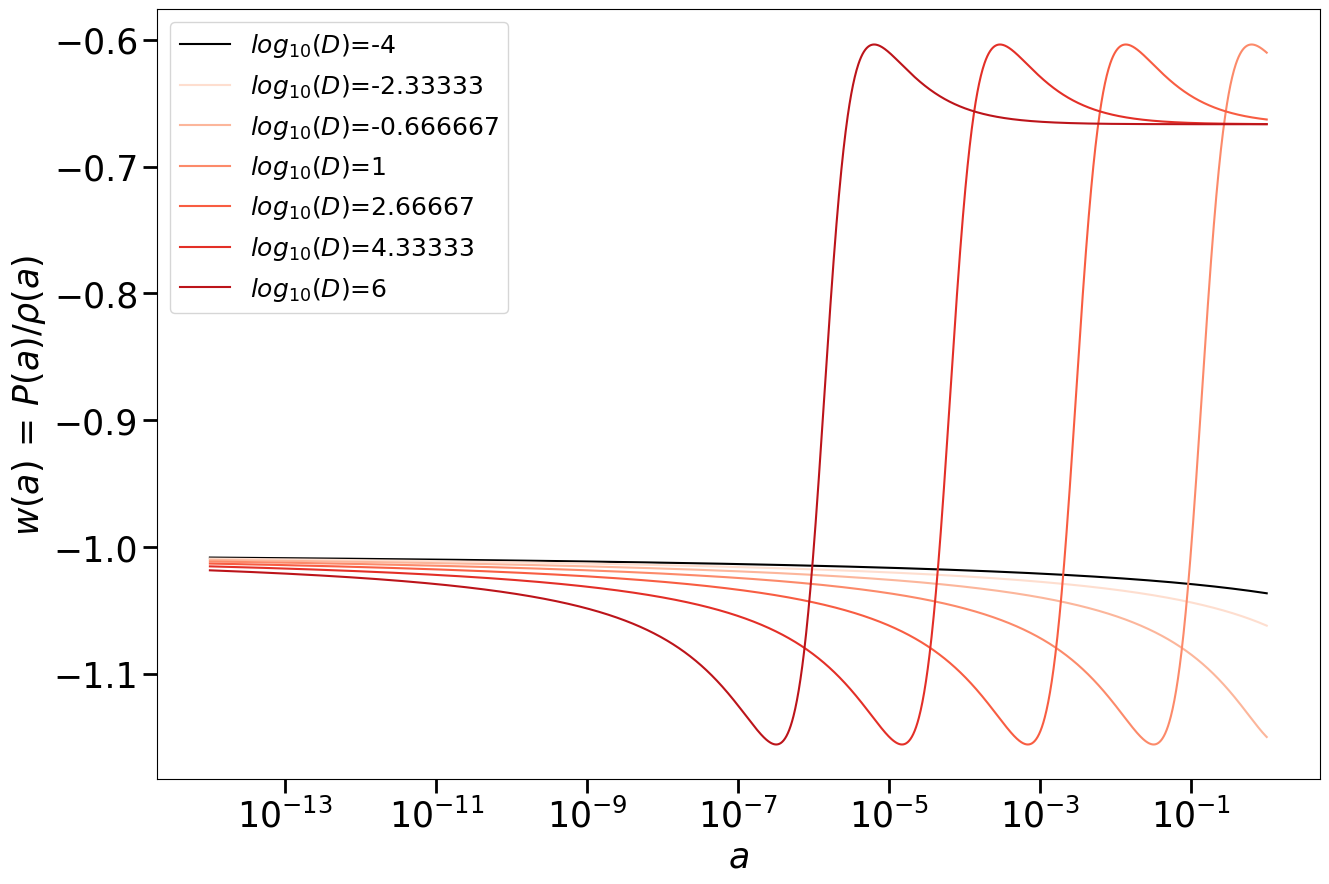}
        \caption {Equation of state evolution for the CG dark energy fluid for different values of its free parameter D.}
    \label{w_CG}
\end{figure}

We turn now to the analysis of the selected model 
defined by the following equation for the dark energy density,
\begin{equation}
\rho_{\rm CG}(a) = \frac{A}{D a - \ln(D a)},
\end{equation}
where the subscript CG stands for CosmoGen.
This model closely resembles viable logarithmic scalar-field dark energy models discussed in the literature (see e.g. \citealt{Wang2023LogScalarDE}).

 In Fig.~\ref{Omega_CG} we show the evolution of the CG energy density, normalised to the $\Omega_\Lambda$ value of the Planck 2018 $\Lambda$CDM model today. Defining the model through its energy density allowed us to write the equation of state 
 as function of the density and its derivatives through Eq.~(\ref{derivw}).  
 Figure~\ref{w_CG} shows 
the evolution of the equation of state of the CG fluid.

The model clearly shows two distinct behaviours. For models with a positive $\log(D)$, the CG density increases earlier, and its equation of state crosses to the phantom regime and returns to less negative values.
In contrast, models with a negative $\log(D)$ have a slow evolution of the equation of state.

\begin{figure}[ht]
        \centering
\includegraphics[width=0.9\hsize]{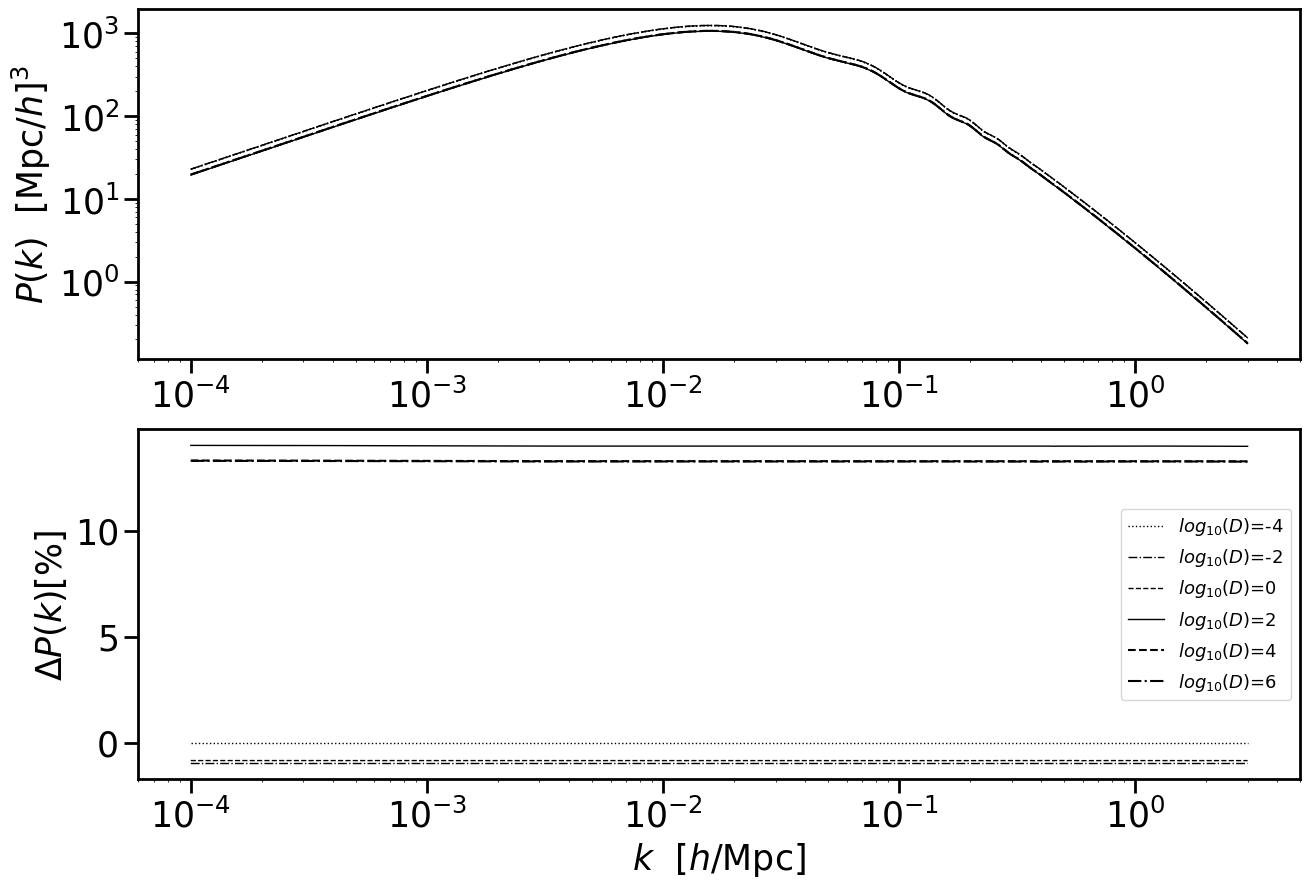}
        \caption {{\em Top panel:} Matter power spectrum for the CG dark energy fluid for different values of D. {\em Bottom panel:} Deviation from $\Lambda$CDM.}
    \label{Pk_CG}
\end{figure}

\begin{figure}[ht]
        \centering
\includegraphics[width=0.9\hsize]{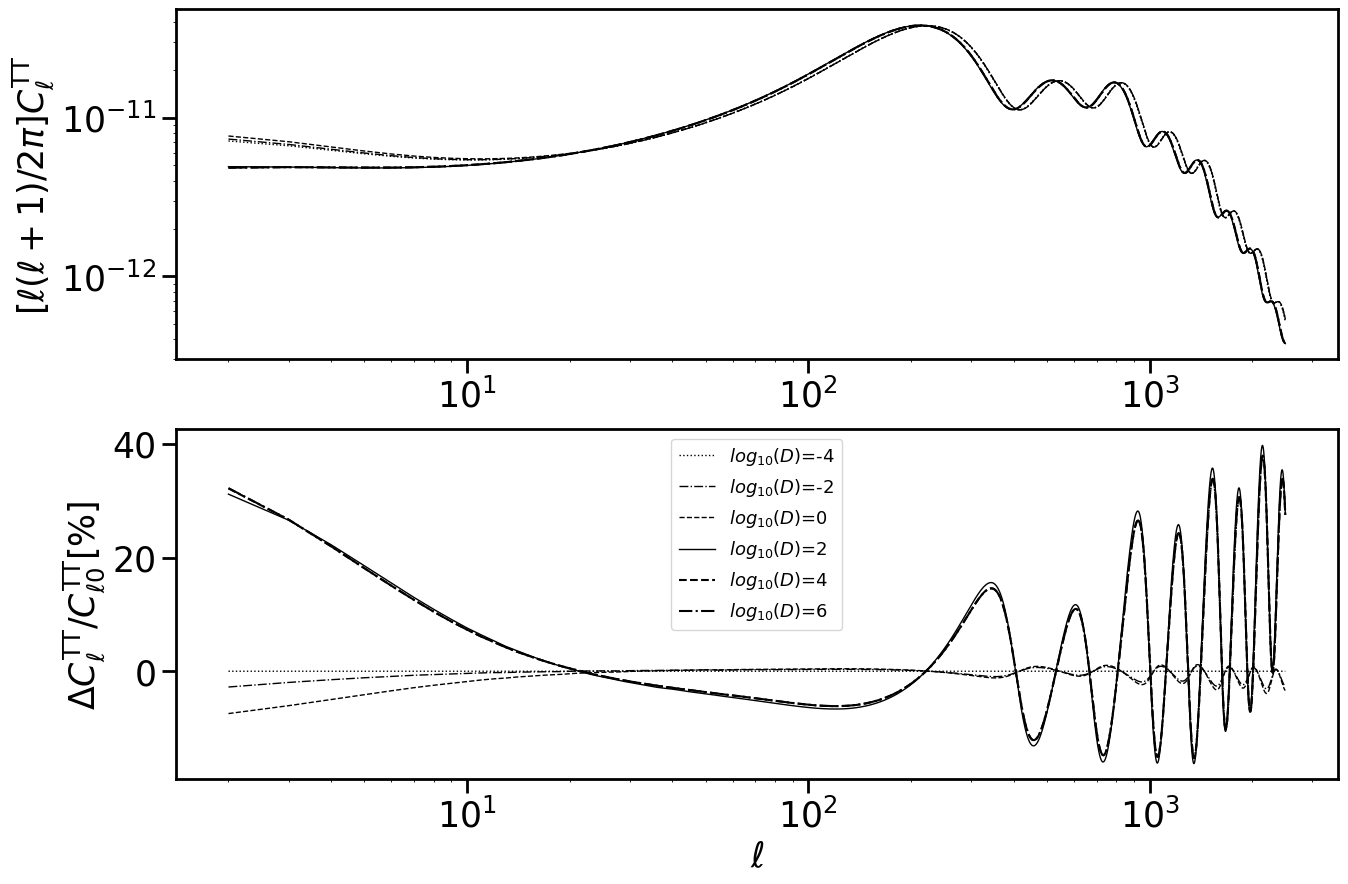}
        \caption {{\em Top panel:} CMB temperature angular power spectrum for the CG dark energy fluid for different values of D. {\em Bottom panel:} Deviation from $\Lambda$CDM.}
    \label{Cl_CG}
\end{figure}

Figures~\ref{Pk_CG} and ~\ref{Cl_CG} show how the model predicted matter power spectrum and CMB temperature angular power spectrum depend on the free parameter $D$. 
Again there are two distinctive behaviours. Negative values of $\log(D)$ can get close to $\Lambda$CDM, while positive values show deviations up to $40\%$ within the probed range.

We performed a likelihood analysis of the model to assess its viability and potential to alleviate cosmological tensions. For this, we no longer used the $H_0$ and $S_8$ priors that were used for model generation. Instead, we tested the model with structure formation data without imposing tension-related constraints. We made an analysis with CMB data (Planck 2018; \citealt{2020A&A...641A...6P}) and another with WL data (KiDS-Viking KV-450; \citealt{2020A&A...633A..69H}) 
We also performed an analysis of $\Lambda$CDM in the same conditions for comparison. 

The parameter space consisted of the CG parameter $D$; the amplitude and slope of the primordial power spectrum, respectively $A_{\rm s}$ and $n_{\rm s}$; the baryon density, $\Omega_{\rm b}$; the CDM density, $\Omega_{\rm CDM}$; and the Hubble constant, $h$. The nuisance parameters we used are the $A_{\rm Planck}$ for CMB and the amplitude of the intrinsic alignment model $A_{\rm IA}$ for WL. We only considered flat models, and other cosmological and nuisance parameters were kept fixed. For all parameters, the analyses were done with flat priors. For $D$, we considered a log(D) range of [-5, 3]. The sampling was made with the PolyChord nested sampling method \citep{Handley2015}, with settings ${\rm PC_{Nlive} = 150}$ (a large number of live points, to increase the accuracy of posteriors and evidence), ${\rm PC_{NumRepeats} = 30}$  (the number of slice-sampling steps to generate a new point, which is relevant for the reliability of the algorithm), and ${\rm PC_{PrecisionCriterium} = 0.001}$  (the end of sampling criterion; sampling terminates when the evidence contained in the live points is a 0.001 fraction of the total evidence).

\begin{figure}[t]
        \centering
\includegraphics[width=0.9\hsize]{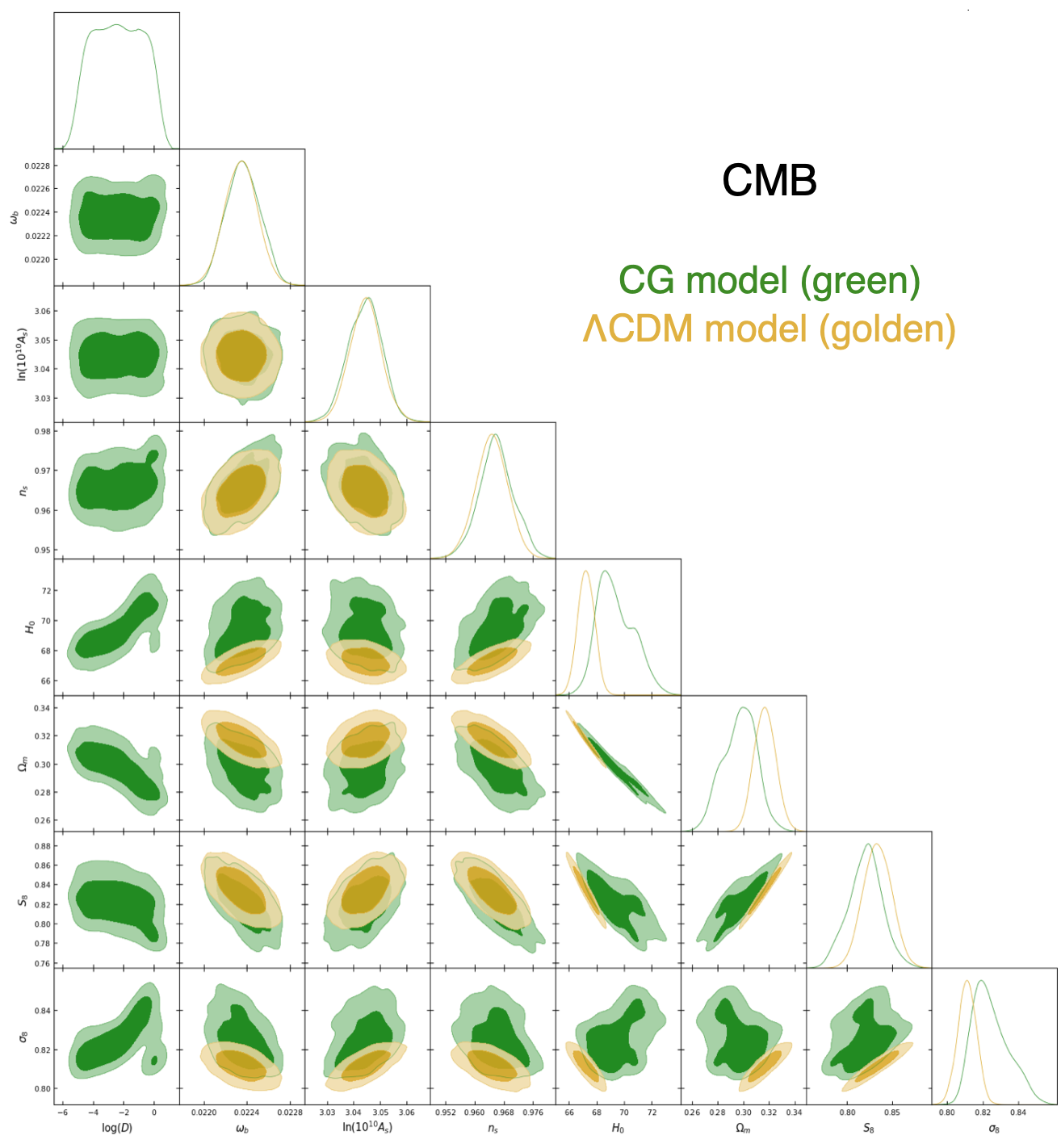}
        \caption {CMB Planck 2018 analyses. Marginalised 2D 1$\sigma$ and 2$\sigma$ contours of the posterior and 1D marginalised posterior 
    for the CG model (green) and $\Lambda$CDM (gold).}
\label{cmbcg}
\end{figure}

\begin{figure}[t]
        \centering
\includegraphics[width=0.9\hsize]{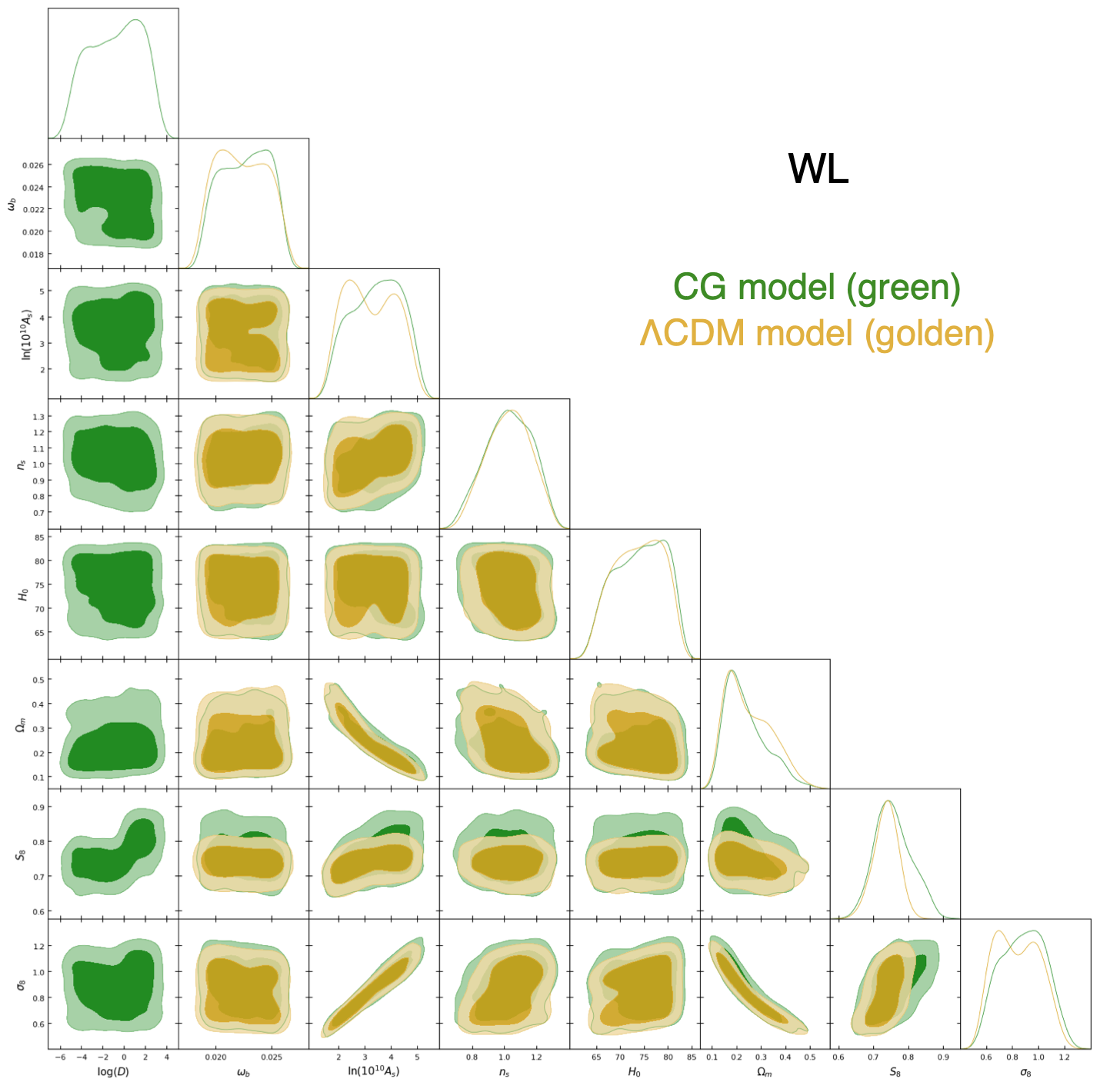}
    \caption {WL KV-450 analyses. Marginalised 2D 1$\sigma$ and 2$\sigma$ contours of the posterior and 1D marginalised posterior 
    for the CG model (green) and $\Lambda$CDM (gold).}
\label{wlcg}
\end{figure}

The resulting constraints for the six basis and three derived parameters of the CG and $\Lambda$CDM models are shown in Table~\ref{tablecg} (indicating the mean and 68\% uncertainty estimated from each dataset) and in Figs.~\ref{cmbcg} and ~\ref{wlcg}. The comparison between the CG model and the $\Lambda$CDM CMB contours in Fig.~\ref{cmbcg} (see also Table~\ref{tablecg}) shows that the CG model pushes $H_0$ to higher values and $\Omega_{\rm m}$ to lower values. The constraints on these two parameters are broader for the CG model, but are comparable for most of the other parameters. The log(D) free parameter is constrained to be negative, as was already hinted at by the model behaviour shown in Figs.~\ref{Omega_CG} to \ref{Cl_CG}. 

\begin{table*}[t]
\centering
\caption{Mean and 68\% uncertainty estimates for the CG and $\Lambda$CDM parameters from the two datasets.}
\begin{tabular} {l l l l l}
\hline
 & \multicolumn{2}{c}{\hspace{-0.9cm}Planck}  & \multicolumn{2}{c}{\hspace{-0.5cm}KiDS} \\[0.1cm]
  Parameter\hspace{0.5cm} & \multicolumn{1}{c}{\hspace{-0.5cm}CG Model}  & \multicolumn{1}{c}{\hspace{-1cm}$\Lambda$CDM}   & \multicolumn{1}{c}{\hspace{-1cm}CG Model}  & \multicolumn{1}{c}{$\Lambda$CDM}
  \\
\hline\\ [-0.3cm]

{\boldmath$\log(D)$} & $-2.3\pm 1.5$ & n.a. & $-0.8^{+3.2}_{-2.5}$ & n.a. \\[0.09cm]

{\boldmath$\omega_{b }$} & $0.02236\pm 0.00015$ & $0.04876\pm 0.00069$ & $0.0226^{+0.0028}_{-0.0020}$ & $0.0401^{+0.0047}_{-0.0081}$ \\[0.09cm]

{\boldmath$\Omega_{c}$} & $0.249^{+0.013}_{-0.010}$ & $0.2582\pm 0.0076$ & $0.196^{+0.044}_{-0.11}$ & $0.207^{+0.085}_{-0.11}$  \\[0.09cm]

{\boldmath$H_{0}$} & $69.4^{+1.3}_{-1.7}$ & $67.88\pm 0.62$ & $74.2^{+7.1}_{-4.3}$ & $73.8^{+6.4}_{-4.6}$ \\ [0.09cm]

{\boldmath$n_{s }$} & $0.9660^{+0.0045}_{-0.0050}$ & $0.9702\pm 0.0043$ & $1.03^{+0.17}_{-0.13}$& $1.02\pm 0.13$ \\[0.09cm]

{\boldmath$\ln 10^{10}A_{s }$} & $3.0443\pm 0.0061$ & $3.1217\pm 0.0058$ & $3.46^{+1.3}_{-0.84}$ & $3.24\pm 0.93$ \\[0.09cm]

{\boldmath$\Omega_{m}$} & $0.297^{+0.015}_{-0.012}$ & $0.3084\pm 0.0083$ & $0.239^{+0.046}_{-0.11}$ & $0.253^{+0.066}_{-0.12}$ \\[0.09cm]

{\boldmath$\sigma_8$} & $0.8246^{+0.0069}_{-0.013}$ & $0.8399\pm 0.0053$ & $0.89\pm 0.16$ &  $0.85^{+0.17}_{-0.22}$ \\[0.09cm]

{\boldmath$S_8$} & $0.820^{+0.020}_{-0.018}$ & $0.852\pm 0.016$ & $0.757^{+0.045}_{-0.062}$ & $0.737^{+0.038}_{-0.031}$ \\[0.1cm]

\hline
{\boldmath$\ln \mathcal B_{CG, \Lambda}$}& \multicolumn{2}{c}{\hspace{-0.9cm}$-1.27$}  & \multicolumn{2}{c}{\hspace{-0.5cm}$-0.99$} \\[0.09cm]
\hline
\end{tabular}
\label{tablecg}
\end{table*}

The Bayes factor, also given in Table~\ref{tablecg}, indicates a weak preference for $\Lambda$CDM. We recall that a positive value of the logarithm of the Bayes factor, $\mathcal{B}_{12}$, between models 1 and 2, has a larger strength of evidence for model 1  \citep{Trotta_2008}.

Regarding the WL analysis, the comparison between the CG model and the $\Lambda$CDM contours in Fig.~\ref{wlcg} (see also Table~\ref{tablecg}) shows broad contours without significant differences between the two models.

\begin{figure}[ht]
        \centering
\includegraphics[width=0.9\hsize]{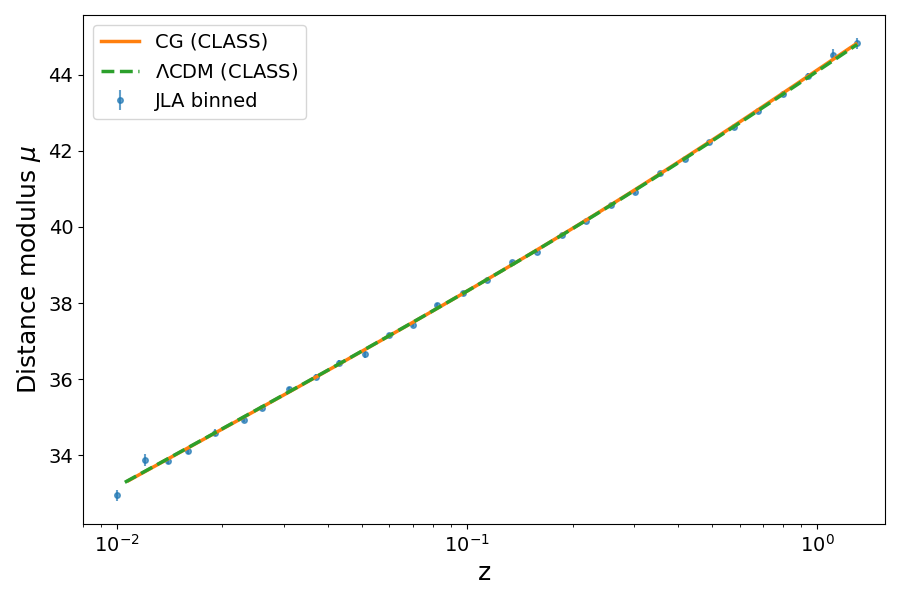}
        \caption {Distance modulus as function of redshift for the GC (orange solid line) and $\Lambda$CDM (green dashed line) models using the Planck estimates of Table~\ref{tablecg}. The data points are from the JLA dataset.}
    \label{SNplot}
\end{figure}

To compare the behaviour of the two models for supernovae observables, we show in 
Fig.\ref{SNplot} the distance modulus computed for the CG and the $\Lambda$CDM models using the parameters obtained in Tab.~\ref{tablecg} (for the CMB analysis). For reference, binned distance modulus measurements from the Joint Light-curve Analysis (JLA) sample \citep{2014A&A...568A..22B} are also plotted. The two curves are very close to each other, with the CG curve deviating from the $\Lambda$CDM one towards higher values on higher redshifts, supporting the findings of a larger Hubble constant.

\begin{figure}[t]
        \centering
\includegraphics[width=0.8\hsize]{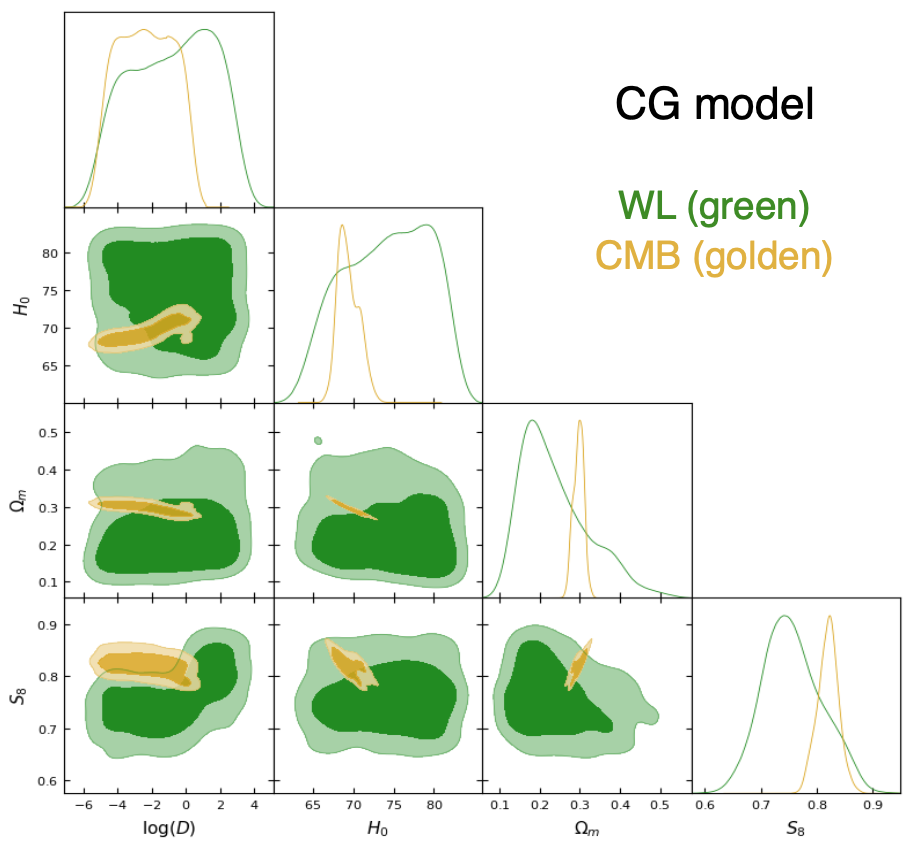}
    \caption {CG model. Marginalised 2D 1$\sigma$ and 2$\sigma$ contours of the posterior and 1D marginalised posterior for the parameters relevant for the Hubble and $S_8$ tensions for CMB Planck 2018 (gold) and WL KV-450 (green).} 
\label{h0cg}
\end{figure}

\begin{figure}[ht]
        \centering
\includegraphics[width=0.8\hsize]{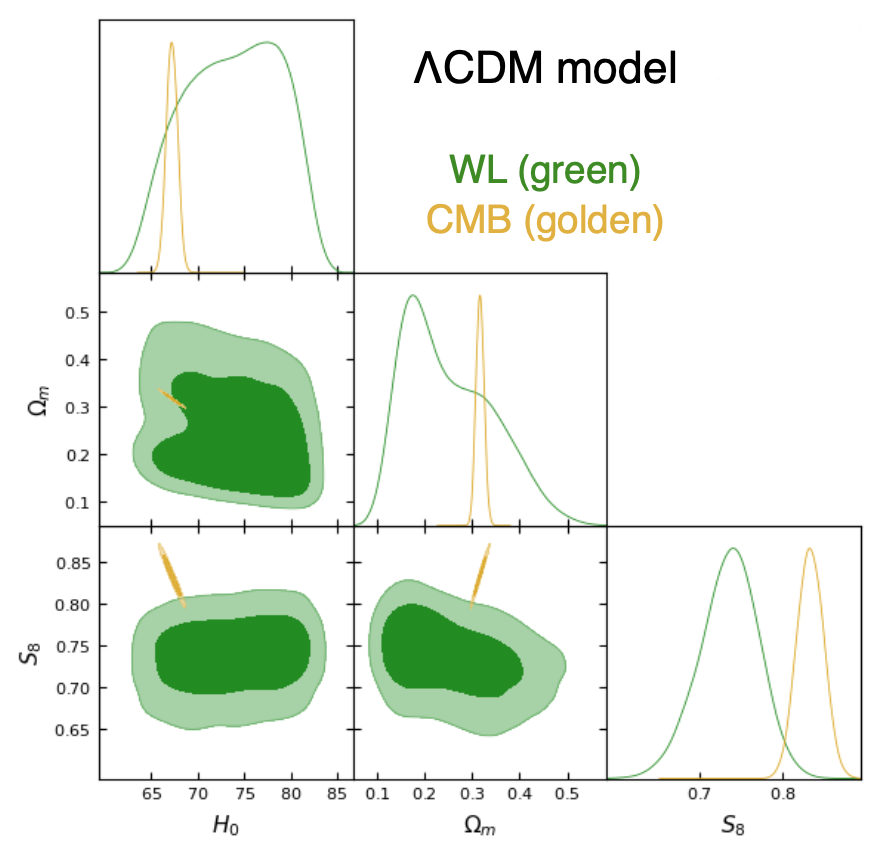}
        \caption {$\Lambda$CDM model. Marginalised 2D 1$\sigma$ and 2$\sigma$ contours of the posterior and 1D marginalised posterior for the parameters relevant for the Hubble and $S_8$ tensions for CMB Planck 2018 (gold) and WL KV-450 (green).}
\label{s8cg}
\end{figure}

The $H_0$ and $S_8$ tensions are better seen in Figs.~\ref{h0cg} and ~\ref{s8cg}, which compare the relevant CMB and WL contours for each of the two models. The plots show a better agreement between the WL and the CMB contours in the case of the CG model. This is especially striking in contours that involve the $S_8$ parameter.

We computed the tension, quantifying the discrepancy between the estimates of the parameter from two different datasets in the standard way in terms of  the combined variances of the two estimates (e.g. \citealt{KiDS-1000}). We find that the CG model shows a $1.1\,\sigma$ $S_8$ tension between the datasets, while for $\Lambda$CDM we found a larger value of $2.8\,\sigma$. Concerning the Hubble tension, we did not quantify it, but we registered a 1$\sigma$ shift of the Planck data $H_0$ CG estimate with respect to the $\Lambda$CDM one towards higher values.

We thus conclude that this case study successfully contributes to the reduction of the tensions. This result is consistent with our approach of intentionally using a likelihood to that effect in the evaluation of the generated models.

\section{Conclusions}
\label{conclusions}

We have presented a new type of cosmological tool, \texttt{CosmoGen}, that integrates traditional methods of numerical cosmology with ML techniques. We showed an example of how \texttt{CosmoGen} can provide new insights into cosmology. Incorporating evolutionary algorithms and likelihood sampling, the tool was applied to generate functional forms of $\rho_{\rm DE}(a)$ that reproduce key features of observational data. Notably, \texttt{CosmoGen} successfully identified fluid dark energy models capable of partially resolving the $S_8$ and $H_0$ tensions.

The potential of \texttt{CosmoGen} is far from fully explored. This tool has the capability to be expanded to generate other types of cosmological models, such as coupled species, unified dark matter, modified gravity theories as well as to map cosmological signatures.

The results we have presented were obtained using a standard laptop with 16 threads. This computational limitation restricted our ability to explore the full range of mathematical functions generated in a single run, leading to lists of candidate models that were often mathematically similar. 
For future development, optimising the code is crucial. The main limitation we encountered was that the stochastic GP method is not the most efficient evolutionary algorithm for sampling the space of possible equations. Further improvements of the code should incorporate alternative methods 
to improve the efficiency of the tool and expand its applicability. There are various other symbolic regression algorithms publicly available, most notably the exhaustive symbolic regression, developed for cosmological applications \citep{10136815}.

On the side of the cosmological likelihood sampling, there is also room for improvement. The option of relying on methods distributed with MontePython led us to use an MCMC method to find the best-fit parameters of the population models. Even though MCMC sampling is more robust in multi-peak distributions, it is 
very costly when compared to local minima finder optimisers. Indeed, the MCMC sampling spends precious time sampling a parameter space and provides much more information than what is strictly needed for model evaluation. In order to reduce computational cost, we imposed further limitations, such as (i) fixing most of the parameters in the evaluation step, (ii) considering dark energy functional forms with only one free parameter, and (iii) limiting the number of generations. These limitations will be addressed in the future development of \texttt{CosmoGen}.

Despite the current limitations in sampling resolution and algorithmic efficiency, the framework demonstrates the potential of integrating AI methods with the broadly used \texttt{CLASS} and \texttt{MontePython} cosmological tools. In this way, \texttt{CosmoGen}, opens a new window for cosmological model building.

\begin{acknowledgements}
We thank the anonymous referee for suggestions that led to an improvement of the paper. We also thank Antonio da Silva for useful discussions and acknowledge support from the Funda\c{c}\~ao
para a Ci\^encia e a Tecnologia (FCT) through R\&D Unit funding UIDB/04434/2020 and UIDP/04434/2020. 
\end{acknowledgements}

\bibliographystyle{aa}
\bibliography{references}

\begin{appendix}

\section{Robustness tests}
\label{tests}

We show the top ranked models from two other runs of \texttt{CosmoGen} in
Tables~\ref{tab:top_equations_2} and \ref{tab:top_equations_3}. 

There are some models in common between the results of the various runs. We notice, however, that models from the same run seem to share some features, while the functional forms are more distinct between different runs.
For example, in the case of Table~\ref{tab:top_equations_2} there is a strong presence of logarithms, while in Table~\ref{tab:top_equations_3} exponentials are more 
. This indicates that a larger initial population is required to accommodate a broad range of functions in a single run. 
 Additionally, this highlights the lower efficiency of GP when compared to other symbolic regression methods, as discussed in \citet{10136815}.

We also note that the $\chi^2_{\rm min}$ of these models are lower than those of Table~\ref{tab:top_equations_1}. The main difference between the setups is that the evaluation MCMC chains used for Table~\ref{tab:top_equations_1} are shorter.
It is also worth mentioning that, in contrast, the $\chi^2_{\rm min}$ values in Tables~\ref{tab:top_equations_2} and ~\ref{tab:top_equations_3} are closer to the $\Lambda$CDM $\chi^2_{\rm min}$ for the same datasets.

It is also interesting to monitor the evolution through the generations. For this, we made new runs with longer chains, this time using the 
PolyChord nested sampling algorithm distributed in \texttt{MontePython} \citep{Handley2015}. An example run is shown in Table~\ref{tab:CG_top_equations_by_gen_split}. In there, we show the best models found in each of seven generations, and observe that the population globally evolves towards more complex and best-fit models.

\begin{table}[ht!]
    \centering
        \centering
    \caption{The 20 best scoring models generated by test run 1.}
        \begin{tabular}{cccc}
            \toprule
            Model & Score & $\chi_{min}^2$ & Complexity \\
            \midrule
            $\frac{A}{a^{\ln{\left(a^{2} \right)}}}$ & 583.17 & 576.17 & 7 \\
            $\frac{A}{\left(D a^{2}\right)^{\ln{\left(a \right)}}}$ & 587.12 & 578.12 & 9 \\
            $\frac{A}{- D + a - \ln{\left(a \right)}}$ & 593.76 & 585.76 & 8 \\
            $\frac{A}{\left(a + e^{- D \ln{\left(a^{2} \right)}}\right)^{D}}$ & 595.20 & 582.20 & 13 \\
            $\frac{A}{\left(D^{2} a\right)^{\ln{\left(a \right)}}}$ & 596.48 & 587.48 & 9 \\
            $\frac{A \left(a^{D}\right)^{D^{2}}}{D}$ & 600.48 & 590.48 & 10 \\
            $\frac{A}{\left(D a\right)^{\ln{\left(a \right)}}}$ & 601.15 & 593.15 & 8 \\
             $\frac{A}{\left(2 a\right)^{\ln{\left(\ln{\left(D^{a} \right)} \right)}}}$ & 601.27 & 591.27 & 10 \\
             $\frac{A}{D^{2} - \ln{\left(a \right)}}$ & 603.04 & 596.04 & 7 \\
             $\frac{A}{- D^{2} + a - \ln{\left(a \right)}}$ & 603.51 & 594.51 & 9 \\
            $A \left(a^{2}\right)^{a}$ & 603.91 & 597.91 & 6 \\
            $A \left(e^{D a^{a}}\right)^{a}$ & 604.06 & 594.06 & 10 \\
            $\frac{A}{D - \ln{\left(a \right)}}$ & 604.87 & 598.87 & 6 \\
            $A e^{a \left(D a\right)^{a}}$ & 604.94 & 594.94 & 10 \\
            $A \left(e^{a^{a + 1}}\right)^{D}$ & 605.96 & 596.96 & 9 \\
             $A \left(e^{D \ln{\left(a^{2} \right)}}\right)^{a}$ & 606.44 & 596.44 & 10 \\
            $A a^{D}$ & 606.59 & 601.59 & 5 \\
            $\frac{A}{- D^{2} a^{D + 1} + a - \ln{\left(a \right)}}$ & 608.02 & 594.02 & 14 \\
            $A e^{a}$ & 609.35 & 605.35 & 4 \\
            \bottomrule
        \end{tabular}
        \label{tab:top_equations_2}
    \end{table}

\begin{table}[ht!]
    \centering
        \centering
         \caption{The 20 best scoring models generated by test run 2.}
        \begin{tabular}{cccc}
            \toprule
            Model & Score & $\chi_{min}^2$ & Complexity \\
            \midrule
            $A a$ & 587.96 & 584.96 & 3 \\
            $A a^{D}$ & 591.81 & 587.81 & 5 \\
            $A e^{D \ln{\left(a \right)}}$ & 591.90 & 584.90 & 7 \\
            $\frac{A}{- D + a - \ln{\left(a \right)}}$ & 592.30 & 583.30 & 8 \\
             $A \left(D a\right)^{2 D} e^{- a}$ & 593.86 & 582.86 & 11 \\
             $A e^{D \ln{\left(2 a \right)}}$ & 594.28 & 586.28 & 8 \\
             $\frac{A}{a^{- D - a} + 1}$ & 595.95 & 586.95 & 9 \\
              $A e^{a - a^{\ln{\left(a^{a} \right)}}}$ & 596.85 & 585.85 & 11 \\
             $-\frac{A}{D \left(a^{a}\right)^{\ln{\left(a \right)}} - D^{a + 1}}$ & 597.19 & 581.19 & 16 \\
             $A \left(2 a\right)^{\left(a^{a}\right)^{a}}$ & 597.20 & 587.20 & 10 \\
             $\frac{A}{\left(e^{a^{2 a \ln{\left(a \right)}}}\right)^{D}}$ & 597.86 & 587.86 & 10 \\
              $\frac{A a^{e^{a^{a}}}}{a^{e^{a^{a}} + 1} + 1}$ & 598.27 & 581.27 & 17 \\
             $A \left(2 a\right)^{\left(a^{a}\right)^{D}}$ & 598.76 & 588.76 & 10 \\
             $\frac{A}{D - \ln{\left(D + a \right)}}$ & 599.09 & 591.09 & 8 \\
             $A e^{a + a^{D} - a^{a}}$ & 599.20 & 587.20 & 12 \\
             $A a^{a^{a} - \ln{\left(a \right)}}$ & 599.95 & 589.95 & 10 \\
            $A e^{a + a^{D} - \left(a^{D}\right)^{a}}$ & 600.46 & 586.46 & 14 \\
            $A \left(2 a\right)^{a^{D}}$ & 601.43 & 593.43 & 8 \\
            $A e^{a}$ & 602.72 & 598.72 & 4 \\
             $\frac{A}{D - a \left(D + a\right) - a + e^{a}}$ & 606.55 & 592.55 & 14 \\
            \bottomrule
        \end{tabular}
        \label{tab:top_equations_3}
\end{table}

\onecolumn

\begin{table}[ht!]
\centering
\caption{The 5 best models of each generation in the Polychords run.}
\label{tab:CG_top_equations_by_gen_split}
\centering
\begin{tabular}{cccccc}
\toprule
Generation & Model & Score & $\chi^2_{min}$ & Complexity & <Complexity>\\
\midrule
 & $\frac{A}{D^{\ln(a)}}$ & 1000.33 & 994.33 & 6& \\
 & $\frac{A}{\ln(Da)}$ & 1016.10 & 1010.10 &6& \\
1 & $\frac{A}{D^a}$ & 1022.98 & 1017.98 &5&6.0\\
 & $\frac{A}{D - \exp(a)}$ & 1043.76 & 1037.76 &6&\\
 & $-\frac{A}{D - \exp(a)}$ & 1045.49 & 1038.49 &7&\\
\midrule
 & $\frac{A}{(D + a)^{\ln(a)}}$ & 950.95 & 942.95 &8&\\
 & $\frac{A}{D^{\ln(2a)}}$ & 1003.96 & 996.96 &7&\\
2 & $A \, D^{-a - 1}$ & 1021.55 & 1016.55 &5&6.2\\
 & $A \, D^a$ & 1021.84 & 1016.84 &5&\\
 & $-\frac{A}{D^a}$ & 1023.00 & 1017.00 &6&\\
\midrule
 & $\frac{A}{(D^a + a)^{\ln(a)}}$ & 953.16 & 943.16 &10&\\
 & $\frac{A}{(a^2 \exp(a))^{\ln(a)}}$ & 965.72 & 955.72 &10&\\
3 & $\frac{A}{D^a (D^a + a^2)}$ & 989.72 & 979.72 &10&9.0\\
 & $\frac{A}{(2D)^{\ln(a)}}$ & 1000.83 & 993.83 &7&\\
 & $\frac{A}{(D(a + 1))^a}$ & 1010.40  & 1002.40 &8&\\
\midrule
 & $\frac{A}{(D a^3)^{\ln(a)}}$ & 890.24 & 881.24 &9&\\
 & $\frac{A}{(D(D - a) + a)^{\ln(a)}}$ & 955.01 & 943.01 &12&\\
4 & $\frac{A \cdot a^{-\ln(a) - 1}}{D^a}$ & 970.26 & 959.26 &11&10.0\\
 & $\frac{A}{a^2 - \ln(a^2)}$ & 973.91 & 965.91 &8&\\
 & $A\, D^{-D^a - \ln(a)}$ & 978.60 & 988.60 &10&\\
\midrule
 & $\frac{A}{(2D a^3)^{\ln(a)}}$ & 887.53 & 877.53 &10&\\
  & $\frac{A}{(D^2 a^3)^{\ln(a)}}$ & 889.29 & 879.29 &12&\\
5 & $\frac{A}{(a^3 \exp(a^D))^{\ln(a)}}$ & 891.02 & 879.02 &10&11.0\\
 & $A \, (2a)^{D - \exp(a)}$ & 911.57 & 903.57 &14&\\
 & $\frac{A}{(D^{a^a + 1} a^2)^{\ln(a)}}$ & 916.64& 902.64 &9&\\
\midrule
 & $\frac{A}{(D a^6)^{\ln(a)}}$ & 846.89 & 837.89 &9&\\
 & $\frac{A}{(D a^3)^{\ln(a^2)}}$ & 848.70 & 838.70 &10&\\
6 & $\frac{A}{(D^2 a^3)^{\ln(a^2)}}$ & 850.92 & 839.92 &11&10.4\\
  & $\frac{A}{(D a^5)^{\ln(a)}}$ & 856.78 & 847.78 &13&\\
 & $\frac{A}{(D a^3)^{\exp(a) \ln(a)}}$ & 859.92& 846.92 &9&\\
\midrule
 & $\frac{A}{(D^2 a^3)^{\ln(a^3)}}$ & 832.30 & 821.30 &11&\\
 & $\frac{A}{(a^3 \exp(a^D))^{\ln(D^{a + 1} a^2)}}$ & 842.40 & 824.40 &18&\\
7 & $\frac{A}{(2D a^6)^{\ln(a)}}$ & 846.38 & 836.38 &15&13.2\\
 & $\frac{A}{(D a^4 \exp(a)^{\exp(a)})^{\ln(a)}}$ & 851.07 & 836.07 &10&\\
 & $\frac{A}{(D^2 a^3)^{\ln(2a^2)}}$ & 852.34 & 840.34 &12&\\
\bottomrule
\end{tabular}
\end{table}

\twocolumn
\end{appendix}

\end{document}